\documentclass[12pt]{article}

\usepackage{amsmath, amssymb, amsthm}
\usepackage{graphicx}

\title{Why Empirical $p$-Values Are Not Uniform:
	Reference Samples, Dependence, and PIT Backtesting}
\author{
	Jakub Lis%
	\thanks{This paper is based on independent research. The views expressed are those of the author alone.\\
	mail: lis.kuba@gmail.com} }

\date{\today}

\begin{document}
	
	\maketitle
	
	\begin{abstract}
	Probability integral transforms (PITs) and empirical $p$-values are widely used to assess the calibration of predictive distributions.
	While exact PIT values are uniformly distributed under correct model specification, practical implementations rely on empirical estimates constructed from finite samples.
	We show that this estimation step fundamentally alters the statistical structure of the problem.
	In particular, common-sample and rolling-window implementations introduce dependence and variance distortions that invalidate classical one-sample uniformity tests.
	When empirical percentiles are conditioned on a shared reference sample, the resulting statistics converge towards a two-sample Kolmogorov--Smirnov regime, while rolling windows induce autocorrelation and variance suppression.
	Our findings indicate that treating empirical percentiles as independent uniform draws can distort statistical inference and that backtesting procedures based on PITs require revised calibration methods accounting for the underlying two-stage sampling structure.
		
	\end{abstract}
	
	\section{Introduction}
	
	Assessing the calibration and stability of predictive distributions is a central task in statistical model validation, with applications ranging from probabilistic forecasting to financial risk management.
	In practice, the probability integral transform (PIT) is widely used as a diagnostic tool for detecting misspecification and distributional change.
	Procedures based on testing the uniformity of empirical percentiles have therefore become common across a range of applied settings, including ensemble forecasting and financial backtesting.
	
	One particularly prominent application arises in financial regulation.
	In particular, the European Central Bank recommends the use of PIT-based procedures in the validation of internal risk models~\cite{ecb}.
	
	The motivation for this approach is commonly formulated as follows (see~\cite{ecb}):
	\begin{quotation}
		If $X$ is a continuous random variable with cumulative distribution function $F_X$, then the transformed variable $F_X(X)$ is uniformly distributed on $[0,1]$.
		Conversely, if the transformed distribution deviates from uniformity, then $F_X$ cannot represent the true distribution of $X$.
	\end{quotation}
	
	This reasoning originates from classical theoretical developments beginning with Rosenblatt's note~\cite{rosenblatt1952} and has been extended in applied statistical literature, including financial econometrics~\cite{berkowitz, diebold} and probabilistic forecasting~\cite{gneiting}.
	Closely related principles also appear in the use of rank histograms for assessing calibration in ensemble forecasting; see, for example,~\cite{hamill}.
	The corresponding methodology has become widely adopted in practical model validation and regulatory backtesting frameworks.
	
	In this note, we highlight a discrepancy between the theoretical foundation of the test and its practical implementation.
	While the theory concerns exact probability integral transforms, practical implementations necessarily rely on empirical estimates based on finite reference samples.
	As a consequence, the quantities subjected to uniformity testing in practice are not uniformly distributed, even under correct model specification.
	This structural mismatch affects the distribution of standard test statistics and calls into question the direct applicability of classical goodness-of-fit procedures in this setting.
	
	The remainder of this paper is organised as follows. 
	Section~\ref{sec:framework} establishes the theoretical framework of the PIT, defines the interpolation methodology used in regulatory practice, and outlines the standard goodness-of-fit testing suite. 
	Section~\ref{sec:singleSample} analyses the distortion introduced by a single fixed reference sample, demonstrating how the test converges to a two-sample problem. 
	In Section~\ref{sec:indepSamples}, we show that the independent-reference framework typical of rank histograms mitigates the systematic impact of estimation error, as the noise associated with finite reference samples averages out rather than contributing coherently through a common conditioning set.
	Section~\ref{sec:rolling} explores the more complex case of rolling-window implementations and the resulting autocorrelation and variance suppression effects. 
	We conclude in Section~\ref{sec:summary}.
	
	\section{Tests and Tools}\label{sec:framework}
	
	\subsection{Test Specification}\label{sec:test}
	
	The proposed testing procedure consists of two main steps.
	First, each new observation is assigned an empirical percentile based on a reference sample.
	Second, the resulting set of $p$-values is tested under the null hypothesis that it follows the uniform distribution on $[0,1]$.
	
	The first component estimates the percentile of an observation $x$ relative to a reference sample $y_1,\dots,y_n$, assuming that $x$ and all $y_i$ are independently drawn from the same probability distribution $f$.
	The exact percentile is given by
	\begin{equation}
		p(x)=\int_{-\infty}^{x} f(u)\,\mathrm du.
	\end{equation}
	
	For $X$ drawn from this distribution, the random variable $p(X)$ is uniformly distributed on $[0,1]$ with cumulative distribution function $F_U$, a property that underlies the standard PIT formulation.
	However, the true distribution is unknown, and this quantity must be estimated empirically.
	
	Let $y_{(1)}<\cdots<y_{(n)}$ denote the ordered reference sample.
	We assign to these observations empirical percentiles $\tilde p_i=i/(n+1)$.
	Intermediate values are then obtained by interpolation.
	For concreteness and simulations, we consider the interpolation methodology prescribed in ECB internal model guidance~\cite{ecb}, under which the empirical percentile estimate is defined as follows:
	
	\begin{equation}\label{eq:ecb}
		\hat{p}(x) = \begin{cases}
			\displaystyle \frac{y_{(k+1)}-x}{y_{(k+1)}-y_{(k)}} \, \tilde{p}_{k} + \frac{x-y_{(k)}}{y_{(k+1)}-y_{(k)}} \, \tilde{p}_{k+1}, & \text{if } y_{(k)} \le x \le y_{(k+1)}, \quad k=1,\ldots,n-1, \\
			\displaystyle \frac{\left(\frac{\tilde{p}_1}{1-\tilde{p}_1}\right)^{\frac{x}{y_{(1)}}}} {1 + \left(\frac{\tilde{p}_1}{1-\tilde{p}_1}\right)^{\frac{x}{y_{(1)}}}}, & \text{if } x < y_{(1)}, \\
			\displaystyle 1 - \frac{\left(\frac{1-\tilde{p}_{n}}{\tilde{p}_n}\right)^{\frac{y_{(n)}}{x}}} {1 + \left(\frac{1-\tilde{p}_n}{\tilde{p}_n}\right)^{\frac{y_{(n)}}{x}}}, & \text{if } x > y_{(n)}.
		\end{cases}
	\end{equation}
	
	The second step consists of testing whether the resulting empirical values $\{\hat p_i\}$ follow a uniform distribution on $[0,1]$, typically using a Kolmogorov--Smirnov (KS) test.
	
	Under the null hypothesis, these values are commonly treated as independent draws from the uniform distribution.
	Rejection of the null is interpreted as evidence that the evaluated observations and reference sample do not originate from the same distribution.
	
\subsection{Test of Uniformity}\label{sec:ks_test}

The empirical cumulative distribution function forms the basis of standard goodness-of-fit procedures.
Among these, the KS test is the conceptually simplest, frequently used and, in our context, analytically tractable.
However, the concerns we highlight are general and arise in any goodness-of-fit test, including the Anderson--Darling and Cramér--von Mises procedures.

The key theoretical result underlying these methods is Donsker's theorem; see~\cite{texbook}.
Since the probability integral transform $p(x)=F(x)$ is strictly increasing, it is natural to express the problem on the unit interval by writing $x=F^{-1}(p)$ and working with $p\in[0,1]$.
In this representation, the underlying distribution is uniform.

Donsker's theorem then implies that
\begin{equation}
	\sqrt{m}\bigl(F_m(p)-p\bigr)\Rightarrow B^0(p),
\end{equation}
where $B^0$ denotes a Brownian bridge.

\begin{figure}[htbp]
	\centering
	\includegraphics[width=0.6\textwidth]{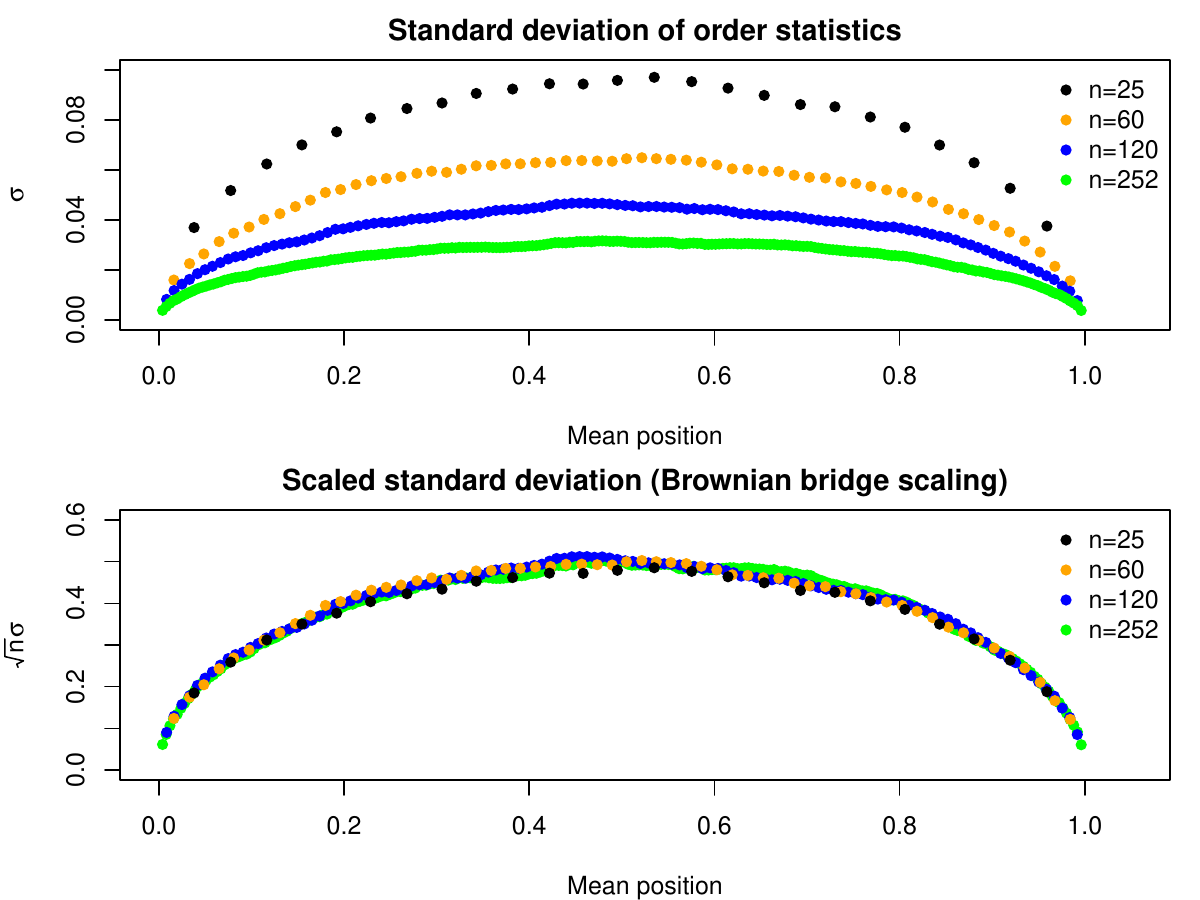}
	\caption{Illustration of Donsker's theorem in finite samples.
		As the sample size increases, the dispersion of the order statistics decreases, consistent with the Brownian bridge approximation.
		After appropriate rescaling, a universal pattern emerges.}
	\label{fig:demo}
\end{figure}

Figure~\ref{fig:demo} illustrates the validity of Donsker's theorem in finite samples and provides a reference for the graphical analyses presented below.
It is based on 1000 Monte Carlo samples of size $n$ drawn from the uniform distribution.
For each replication, we compute the order statistics of the sample.
For each fixed rank $i$, we then compute the empirical mean and standard deviation of $y_{(i)}$ across replications.
As the sample size $n$ increases, the standard deviation of $y_{(i)}$ converges to zero, reflecting the vanishing variability of order statistics in large samples.
However, after rescaling by $\sqrt{n}$, the variance profile across $i$ converges to that of a Brownian bridge, that is, proportional to $t(1-t)$ for $t = i/(n+1)$.
In particular, fluctuations are smallest near the boundaries and largest in the centre of the distribution, reflecting the characteristic shape of the Gaussian bridge.

This result underlies most classical goodness-of-fit procedures, in particular the KS test, which is based on the maximal deviation between the empirical and reference distribution functions.
In the present setting, this corresponds to
\[
\sup_{p \in [0,1]} \left| F_n(p) - p \right|.
\]
Under the null hypothesis, Donsker's theorem implies that
\[
\sqrt{n}\,\sup_{p \in [0,1]} \left| F_n(p) - p \right|
\;\Rightarrow\;
\sup_{p \in [0,1]} |B^0(p)|.
\]
Equivalently, the KS statistic is of order $n^{-1/2}$ in probability.

The KS test can be formulated in two standard settings: the one-sample and the two-sample case.
In the one-sample setting, an empirical distribution is compared with a fully specified reference distribution.
The two-sample setting compares the empirical distributions of two independent samples of size $m$ and $n$, both assumed to be drawn from the same underlying distribution.
In this case, the joint fluctuation scales as $(m^{-1}+n^{-1})^{1/2}$ and can be interpreted as the difference of two independent empirical processes.

\section{Single Conditioning Sample}\label{sec:singleSample}

We now examine the effect of replacing exact percentiles $p(x)$ with empirical estimates $\hat p(x)$ constructed from a finite reference sample $y_1,\dots,y_n$.

Let $y_{(1)} < \cdots < y_{(n)}$ denote the order statistics of the reference sample, and let $F$ be the true cumulative distribution function. 
For an observation $x$ such that $x \in (y_{(i-1)}, y_{(i)})$, the true probability satisfies
\begin{equation}
	p(y_{(i-1)}) < p(x) < p(y_{(i)}).
\end{equation}
Thus, the probability mass of this interval is
\begin{equation}
	p(y_{(i)}) - p(y_{(i-1)}).
\end{equation}

\begin{figure}[htbp]
	\centering
	\includegraphics[width=0.6\textwidth]{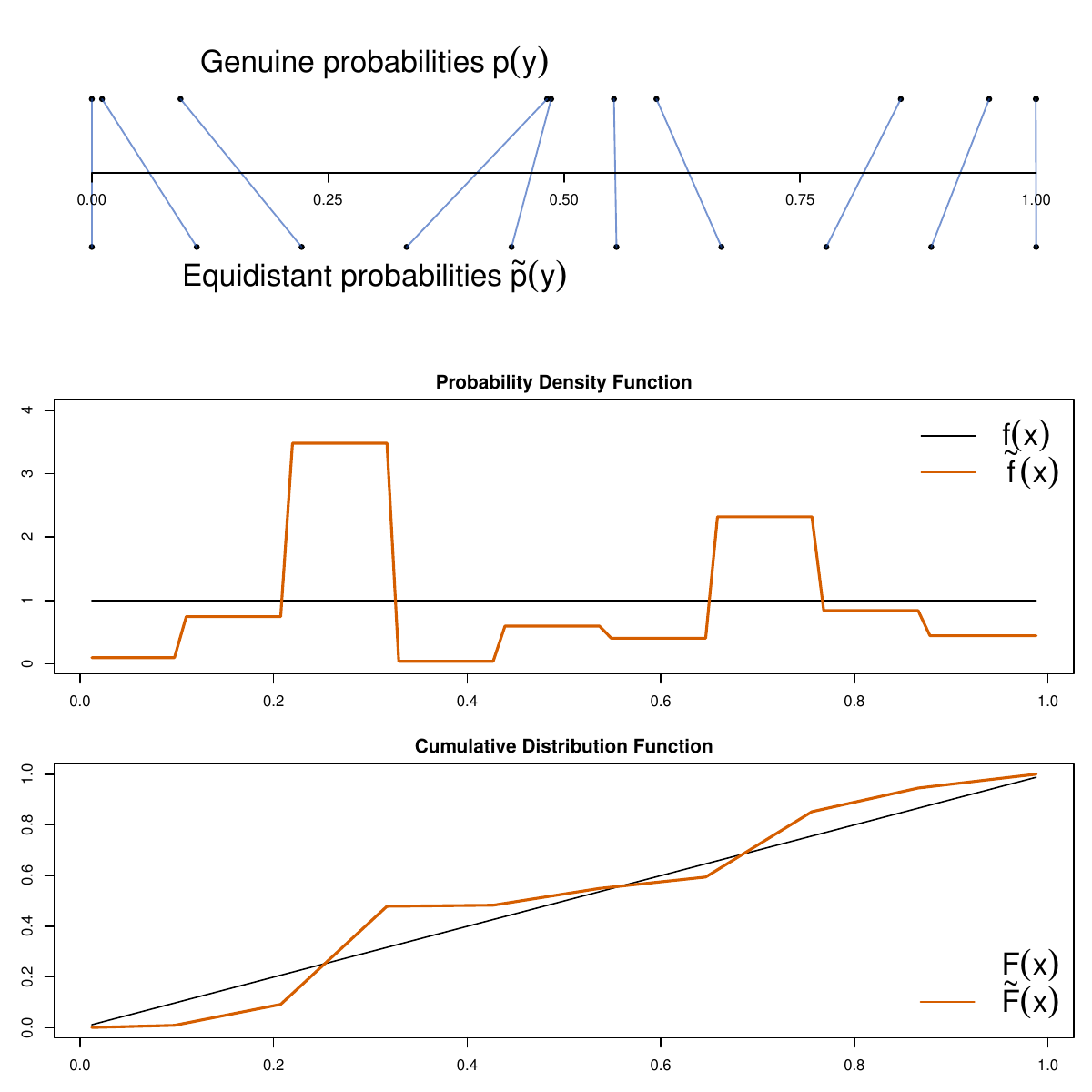}
	\caption{
		Illustration of the distortion induced by empirical probability integral transform estimation.
		The top panel shows the mapping from genuine probabilities $p(y)$ to empirical probabilities $\tilde p(y)$.
		The middle and bottom panels show the resulting deformation of the density and cumulative distribution functions, comparing the distribution of true probabilities $p(x)$ with that of empirical probabilities $\hat p(x)$.
	}
	\label{fig:tildeF}
\end{figure}

In contrast, the empirical construction assigns equal mass $1/(n+1)$ to each interval, so that
\begin{equation}
	\frac{i-1}{n+1} < \hat p(x) < \frac{i}{n+1}.
\end{equation}
These quantities do not coincide in general. This relationship holds irrespective of the specific interpolation scheme used between order statistics, as it reflects the equal-mass structure imposed by the empirical distribution function rather than any particular interpolation choice. 
The construction is illustrated schematically in Figure~\ref{fig:tildeF}.

As a consequence, $\hat p(x)$ does not evaluate the true cumulative distribution function $F(x)$, but rather a sample-dependent approximation thereof. This induces a random distribution, which we denote by $\tilde F$, reflecting the variability of the reference sample.

Given a realisation of the reference sample, the mapping $x \mapsto \hat p(x)$ becomes deterministic. 
Since the $x_i$ are independent, the transformed values $\{\hat p_i\}$ form an i.i.d.\ sample from the distribution induced by $\tilde F$.

This leads to the following decomposition of the empirical distribution function $F_m$ of the values $\{\hat p_i\}$ relative to the uniform distribution $F_U$:
\begin{equation}\label{eq:decomp}
	F_m - F_U
	=
	(F_m - \tilde F)
	+
	(\tilde F - F_U).
\end{equation}

The difference $F_m - F_U$ forms the basis of standard uniformity tests. 
The first term in Eq.~(\ref{eq:decomp}), $F_m - \tilde F$, represents the usual empirical fluctuation of a sample of size $m$ drawn from $\tilde F$, and is of order $\mathcal{O}_p(m^{-1/2})$.
The second term, $\tilde F - F_U$, captures the deviation of $\tilde F$ from the uniform distribution and is of order $\mathcal{O}_p(n^{-1/2})$, reflecting the finite size of the reference sample.

Thus, the empirical process combines two sources of randomness: sampling variability from the evaluated observations and estimation error from the reference sample. 
These contributions are of comparable magnitude unless $n$ is much larger than $m$.

\begin{figure}[htbp]
	\centering
	\includegraphics[width=0.95\textwidth]{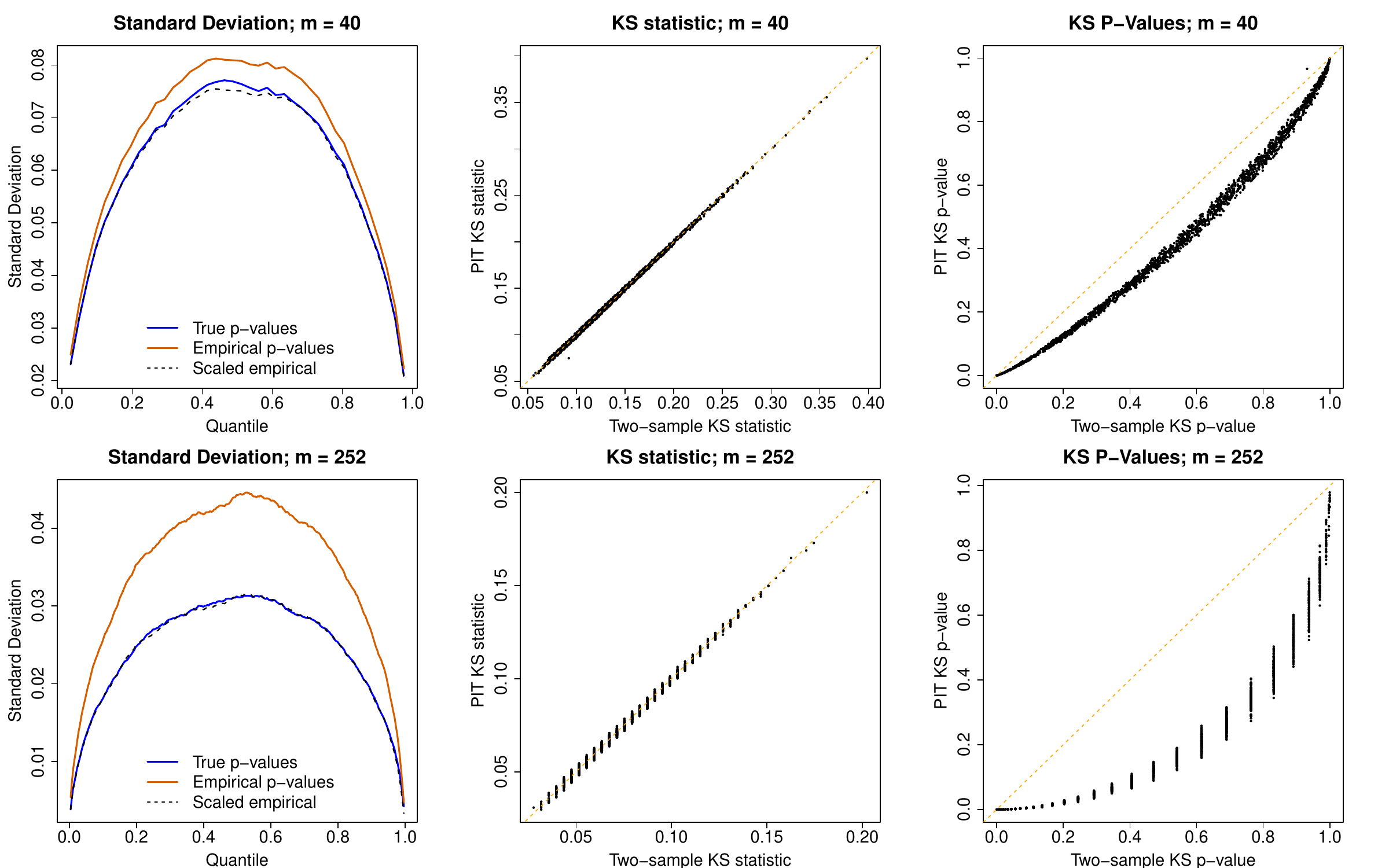}
	\caption{
		Results from 2000 Monte Carlo experiments with 252 observations in the reference sample and the indicated length of the empirical sample ${\hat p_i}$, with observations drawn from a normal distribution.
		Left column: dispersion of exact and empirical percentiles, showing systematic inflation for empirical $\{\hat p_i\}$.
		The black curve corresponds to the standard deviations of empirical percentiles scaled by $\sqrt{1/m}/\sqrt{1/m+1/n}$.
		Middle column: comparison of one-sample KS statistics with corresponding two-sample KS statistics, showing near-perfect agreement.
		Right column: comparison of resulting $p$-values, revealing systematic discrepancies and increased rejection under one-sample calibration.
	}
	\label{fig:twoSample}
\end{figure}

The decomposition in Eq.~(\ref{eq:decomp}) is supported by numerical evidence; see Figure~\ref{fig:twoSample}. 
In particular, the left panel indicates that the fluctuation scale is well approximated by $(m^{-1} + n^{-1})^{1/2}$, consistent with a superposition of two empirical processes. 

Indeed, numerical experiments (see the middle panel of Figure~\ref{fig:twoSample}) show that the KS statistics computed from the empirical $\{\hat p_i\}$ are nearly indistinguishable from those arising in the two-sample case.

This close correspondence for the KS statistic is notable and can be justified analytically; see~\ref{app:pit}.  
It highlights the impact of the two sources of randomness on the distribution of $\{\hat p_i\}$. 
In particular, the set $\{\hat p_i\}$ cannot be regarded as an i.i.d.\ sample from a fixed distribution, and the classical one-sample goodness-of-fit framework is therefore no longer valid.

\section{Independent Reference Samples}\label{sec:indepSamples}

The framework of rank histograms in ensemble forecasting~\cite{hamill} suggests a different relationship between the reference samples and the associated empirical percentiles.
Rather than evaluating all observations against a common conditioning sample, a distinct reference sample is generated for each percentile estimation.
More precisely, suppose that for each observation $x_i$ an independent reference sample
\[
y_{i1},\dots,y_{in}
\]
is generated from a distribution $F_i$, where the distribution itself may vary with $i$.
The only requirement is that, for each fixed $i$, the observation $x_i$ and its associated reference sample are drawn from the same distribution $F_i$.
Thus, unlike in the fixed-reference setting considered above, the distributions underlying different observations need not coincide.

Let $\hat p_i$ denote the empirical PIT value constructed from $x_i$ and its associated reference sample.
In contrast to the fixed-reference framework, the empirical mapping
\[
x_i \mapsto \hat p_i
\]
now changes independently from one observation to the next.

\begin{figure}[htbp]
	\centering
	\includegraphics[width=0.95\textwidth]{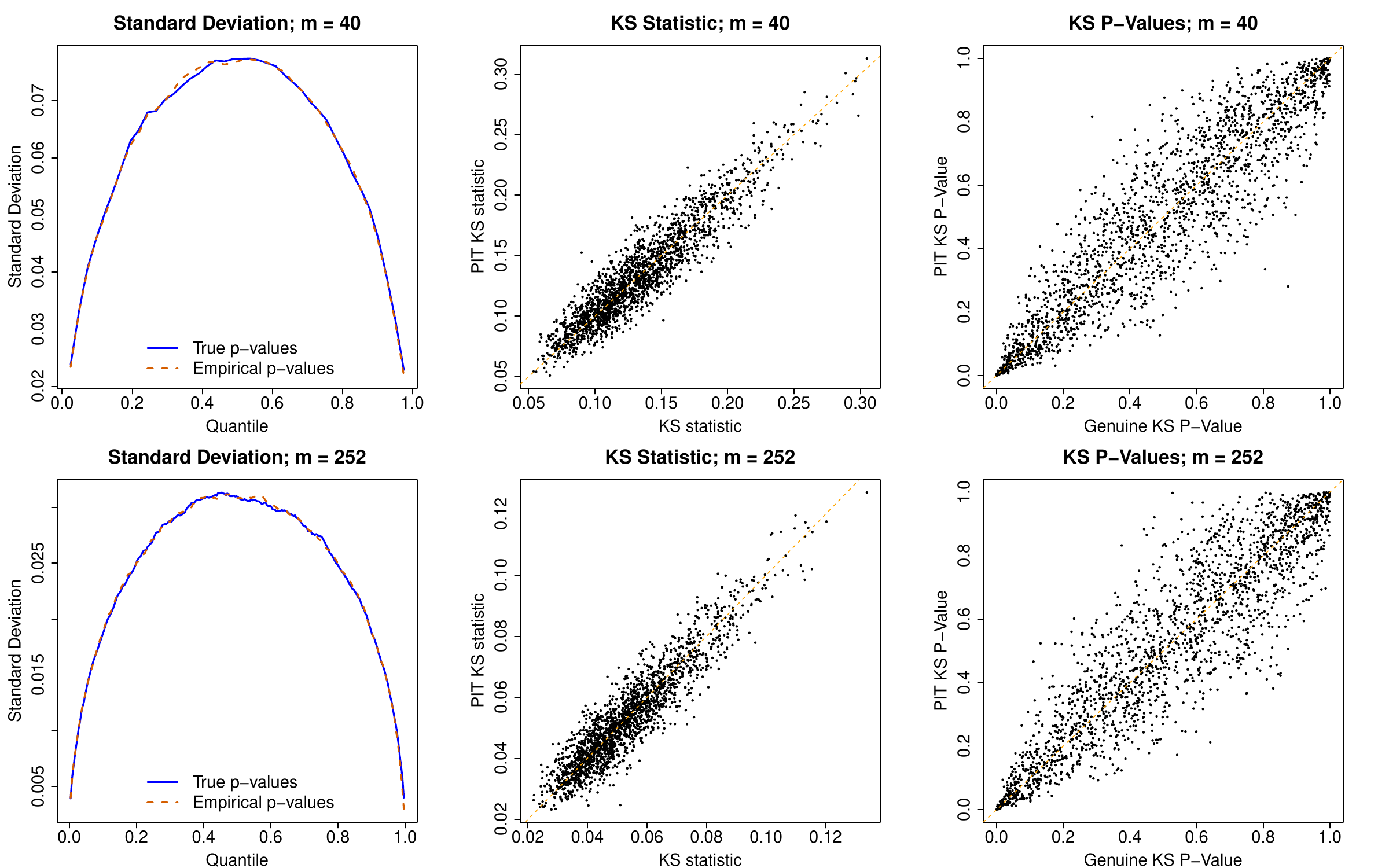}
	\caption{
		Results from 2000 Monte Carlo experiments with 252 observations in the reference samples and the indicated length of the empirical sample $\{\hat p_i\}$, with observations drawn from a normal distribution.
		Left column: dispersion of exact and empirical percentiles $\{\hat p_i\}$, showing close agreement between the two constructions.
		Middle column: comparison of one-sample KS statistics based on exact percentiles and KS statistics based on empirical percentiles, with clear correlation between these two.
		Right column: comparison of the resulting $p$-values. 
		The empirical $p$-values retain information about the underlying distribution.
	}
	\label{fig:manySample}
\end{figure}

Numerical investigations (see Figure~\ref{fig:manySample}) indicate that in this setting the additional variability induced by the empirical reference samples vanishes.
At the level of the decomposition, Eq.~(\ref{eq:decomp}) this reflects the fact that the fluctuations of the random reference distributions $\tilde F_i$ average out across observations, rather than contributing coherently through a common conditioning sample.

This effect is illustrated in the left panels of Figure~\ref{fig:manySample}, where the dispersion of the empirical percentiles $\hat p_{(i)}$ is seen to closely track that of the exact percentiles $p_{(i)}$.

As a particular case, suppose that all observations are drawn from the same underlying distribution.
We then investigate whether the empirical values $\hat p$ retain information about the original sample by comparing the one-sample KS statistics and associated $p$-values computed from the exact and empirical percentiles; see the middle and right panels of Figure~\ref{fig:manySample}.
While the noise induced by the finite reference samples affects both the KS statistics and their associated $p$-values, the relationship to the corresponding exact quantities remains evident.

This discussion naturally points to a potential source of error in statistical procedures: if successive reference samples are not independent, the terms $\tilde F_i $ may fail to average out.
As a consequence, persistent dependence in the estimation noise may systematically affect the distribution of the transformed values $\hat p_i$, thereby distorting subsequent statistical inference.

\section{Rolling Window Framework}\label{sec:rolling}

Another practically relevant implementation computes empirical $p$-values using a rolling reference sample.
In this framework, each observation is evaluated relative to the preceding $n$ observations, after which the reference window is shifted forward by one step.

This construction introduces additional complications.
In particular, substantial dependence is induced between successive $\hat p$-values: each observation determines its own $\hat p$-value and subsequently enters the reference sample used for future evaluations.
As a result, the sequence $\{\hat p_i\}$ can no longer be interpreted as an independent sample.

\begin{figure}[htbp]
	\centering
	\includegraphics[width=0.95\textwidth]{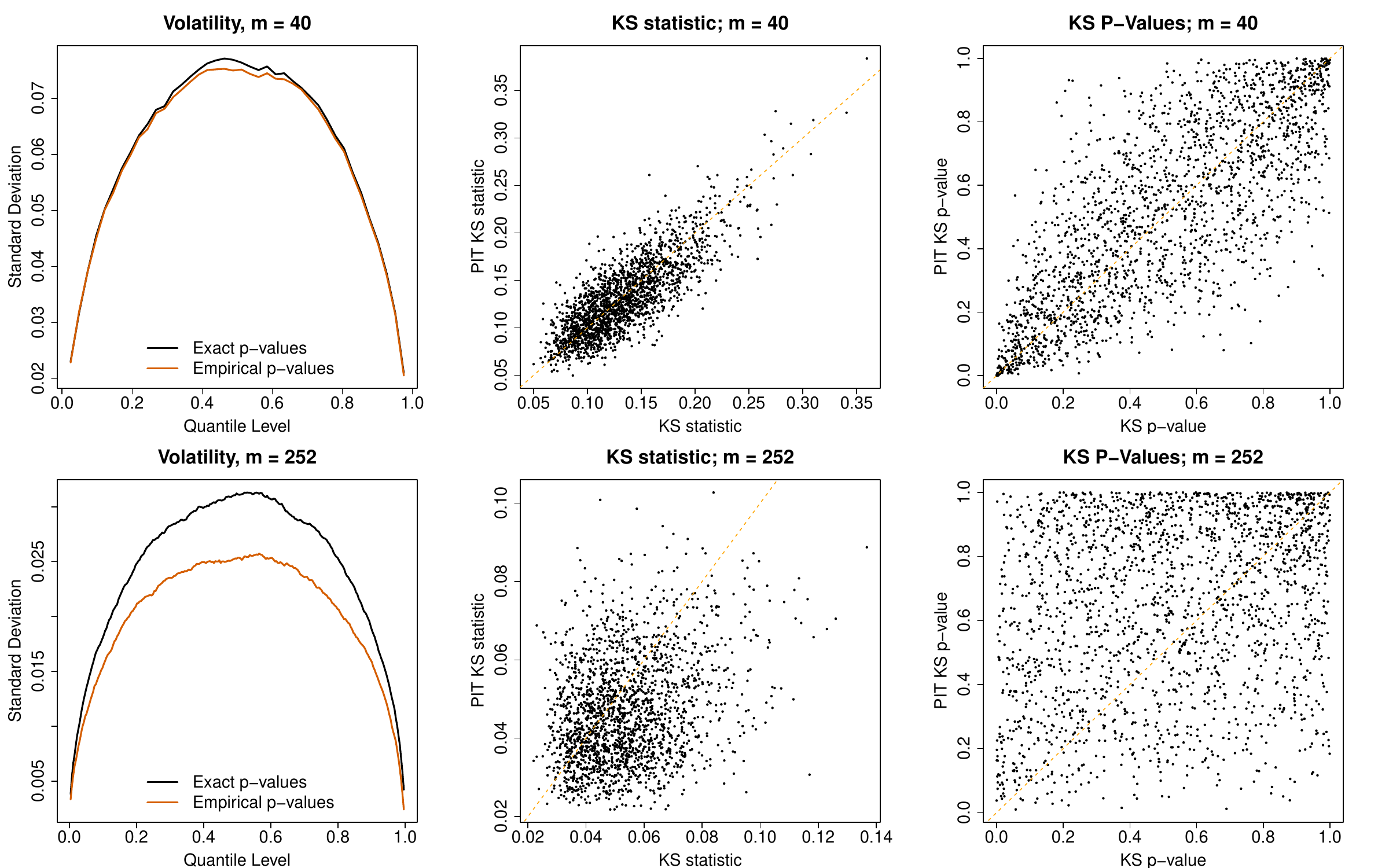}
	\caption{
		Results from 2000 Monte Carlo experiments.
		In each run, a rolling window of size 252 is used with observations drawn from the standard normal distribution, followed by an independent sample of size $m$.
		Left column: dispersion of exact and empirical percentiles in the rolling-window setting, showing a systematic reduction in the variability of empirical $\hat{p}$-values.
		Middle column: comparison of KS statistics computed from exact $p$-values and empirical $\hat{p}$-values.
		Right column: comparison of the corresponding KS $p$-values, showing a progressive loss of correspondence as $m$ approaches $n$.
	}
	\label{fig:autocorel}
\end{figure}

This effect is evident in numerical simulations; see Figure~\ref{fig:autocorel}.
The left panels show that the dispersion of rolling-window order statistics is systematically reduced relative to that of exact $p$-values.
This effect is weak when the number of evaluated $\{\hat p_i\}$ values is small relative to the window size $n$, but becomes pronounced when the two are of comparable magnitude.

The middle and right panels compare KS statistics and corresponding $p$-values computed from exact $\{p_i\}$ and empirical $\{\hat p_i\}$ samples.
When $m \ll n$, fluctuations are dominated by the smaller sample, and a clear correlation between the two statistics is observed.
In this regime, the effect of the conditioning sample does not yet significantly distort the dependence structure.

However, when $m$ and $n$ are of comparable magnitude, this relationship breaks down, and the KS statistics (and their associated $p$-values) become effectively uncorrelated.
This loss of correspondence indicates that empirical $\hat p$-values do not preserve the same distributional behaviour as the true $p$-values.

As a consequence, conclusions drawn directly from $\hat p$-values, as if they were genuine $p$-values, can be misleading.

	\section{Conclusion}\label{sec:summary}
	
	Empirical $p$-values reflect two sources of randomness, thereby invalidating the classical calibration of uniformity tests. 
	In this note, we have shown that when empirical $p$-values are computed using a single reference sample, the resulting KS statistic is close to that of the two-sample setting.
	Consequently, using $p$-values derived from the one-sample KS test can distort statistical inference.
	
	In contrast, independent reference samples mitigate the effect of conditioning noise through averaging, so that the transformed empirical percentiles continue to reflect the statistical properties of the underlying sample.
	
	Rolling-window implementations further increase the complexity of the problem, as the evolving reference sample couples successive empirical $\hat p$-values through overlapping information sets. 
	Developing an analytical understanding of this setting appears more challenging.
	
	Nevertheless, our findings indicate that standard backtesting procedures based on empirical $p$-values require revised calibration methods that account for the two-stage sampling structure.
	
	\appendix
	\renewcommand{\thesection}{Appendix \Alph{section}}
	
	\section{From Two-Sample to PIT-Based KS Statistics} \label{app:pit}
	
	\paragraph{1. Two-sample KS statistic.}
	
	Let $\{x_i\}_{i=1}^m$ and $\{y_j\}_{j=1}^n$ be independent samples with empirical distribution functions $F_m$ and $G_n$. The two-sample Kolmogorov--Smirnov statistic is defined as
	\[
	D_{m,n} = \sup_{t \in \mathbb{R}} |F_m(t) - G_n(t)|.
	\]
	Since both $F_m$ and $G_n$ are step functions with jumps of size $1/m$ and $1/n$, respectively, the supremum is attained at points in the pooled sample $\{x_i\} \cup \{y_j\}$. Thus, the supremum reduces to a maximum over at most $m+n$ candidate points,
	\begin{equation}\label{stat:twosample}
		D_{m,n} = \max_{z \in \{x_i\} \cup \{y_j\}} |F_m(z) - G_n(z)|.
	\end{equation}

	\paragraph{2. One-sample PIT statistic.}
	
	Consider the empirical PIT values $\{\hat{p}_i\}_{i=1}^m$ and their order statistics $\hat{p}_{(i)}$. The one-sample KS statistic against the uniform distribution is given by
	\[
	D_m = \max_{1 \le i \le m} \left| \hat{p}_{(i)} - \frac{i}{m} \right|.
	\]
	
	By construction, for each $x_{(i)}$,
	\[
	\frac{\#\{y_j < x_{(i)}\}}{n+1}
	\;\le\;
	\hat{p}_{(i)}
	\;<\;
	\frac{\#\{y_j < x_{(i)}\} + 1}{n+1},
	\]
	which implies
	\[
	\hat{p}_{(i)} = G_n(x_{(i)}) + \mathcal{O}(n^{-1}).
	\]
	
	Therefore,
	\[
	D_m = \max_{1 \le i \le m} \left| G_n(x_{(i)}) - \frac{i}{m} \right| + \mathcal{O}(n^{-1}).
	\]
	
	Since $F_m(x_{(i)}) = i/m$, 
	\begin{equation}\label{stat:one}
		D_m = \max_{1 \le i \le m} \left| F_m(x_{(i)}) - G_n(x_{(i)}) \right| + \mathcal{O}(n^{-1}),
	\end{equation}
	which corresponds to evaluating the two-sample statistic $D_{m,n}$ only at the $x$-grid points $\{x_{(i)}\}$, rather than over the full pooled set $\{x_i\} \cup \{y_j\}$ used in $D_{m,n}$.
	
	Since the one-sample statistic explores only a subset of the candidate points available in the two-sample case, it cannot attain a larger maximum. Hence,
	\begin{equation}
		D_m \le D_{m,n} + \mathcal{O}(n^{-1}).
	\end{equation}
	
	\paragraph{3. Maximal difference between $D_m$ and $D_{m,n}$.}
	
	Assume that the two-sample maximum is attained at a point $t^*$ lying in an interval $(x_{(i)}, x_{(i+1)})$. On this interval, $F_m$ is constant and $G_n$ is non-decreasing. Hence $F_m - G_n$ is non-increasing.
	
	Two cases arise:
	
	\begin{itemize}
		\item If $F_m - G_n \ge 0$ at the maximiser, then the maximum over the interval is attained at the left boundary $x_{(i)}$, and is therefore included in the $x$-grid. In this case, no loss occurs when passing to $D_m$.
		
		\item If $F_m - G_n < 0$ at the maximiser, then the extremum is attained at the largest $y_{(j)}$ such that $x_{(i)} < y_{(j)} < x_{(i+1)}$. Evaluating the difference at $x_{(i+1)}$ yields
		\[
		F_m(x_{(i+1)}) - G_n(x_{(i+1)}) = \bigl(F_m(t^*) - G_n(t^*)\bigr) + \frac{1}{m}.
		\]
		Thus, the value captured on the $x$-grid differs from the true extremum by at most one jump of $F_m$, that is, $1/m$.
	\end{itemize}
	
	Therefore, restricting the supremum to the $x$-grid introduces a discrepancy of at most $1/m$ relative to the two-sample statistic.
	
	\paragraph{4. Conclusion.}
	
	Two sources of discrepancy arise when comparing $D_m$ and $D_{m,n}$. The first, of order $1/n$, is due to the discretisation inherent in the definition of empirical PIT values. The second, of order $1/m$, results from restricting the supremum to the $x$-grid.
	
	Consequently, the difference between the two statistics is of order $\mathcal{O}(m^{-1} + n^{-1})$. Since the stochastic fluctuations of the empirical process scale as $(m^{-1} + n^{-1})^{1/2}$, this discrepancy is asymptotically negligible, which explains the close agreement observed in numerical experiments. Moreover, this result confirms the decomposition in Eq.~(\ref{eq:decomp}), showing that the variability of empirical percentiles reflects also the randomness of the reference sample.

\end{document}